# VORTEX METROLOGY USING CIRCULAR AND ELLIPTICAL LAGUERRE GAUSS KERNELS

**Héctor Rabal[1,3], Myrian Tebaldi[1,2], Astrid Villamizar Amado[1,4], Silvana Gallo[1,5], Nelly Cap[1,6]**

1. Centro de Investigaciones Ópticas (CONICET La Plata-CIC-UNLP).

2. Departamento de Ciencias Básicas, Facultad de Ingeniería, Universidad Nacional de La Plata, Argentina.

3. UIDET OPTIMO, Departamento de Ciencias Básicas, Facultad de Ingeniería, Universidad Nacional de La Plata, Argentina.

4. Facultad de Ciencias Exactas, Universidad Nacional de La Plata, Argentina.

5. Facultad de Informática, Universidad Nacional de La Plata, Argentina.

6. Comisión de Investigaciones Científicas de la Provincia de Buenos Aires, La Plata, Argentina.

## Abstract

We propose a generalization of the Laguerre Gauss Transform to include elliptical kernels in vortices metrology. To that end, we broaden the original pseudo field obtained using Laguerre Gauss Transform (LGT) that usually includes circularly a symmetric Gaussian core to include two parameters, the widths of the Gaussian filter in two perpendicular directions, thus generating a family of slightly elliptical kernels. This permits the generation of a family of very close vortices that improves the identification of homologous ones and the precision and locality of the measurements. We show examples with laser generated speckle and in ultrasound images.

Corresponding author: h_rabal_geb@yahoo.com

## Introduction

Vortices, it is the existences of loci of a wave field where the field vanishes and the phase is not defined, were first coined by Berry et al [1]. In mathematics, a function is called analytic if it can be expressed as a convergent power series of its variables. There is a theorem (the unicity theorem) in Calculus stating that analytic functions can only be zero in region of null measure or it is zero everywhere. In familiar terms, this means that vortices are, from a mathematical point of view, arbitrarily small. This feature in practice, that can only be approximately attended in practice, make vortices appropriate tools to precisely locate points on a speckle field [2, 3] and to measure low activity in dynamic speckle patterns [4].

We propose a generalization of the Laguerre Gauss Transform (LGT) to help in vortices metrology in both laser and ultrasound speckle.

There exist in the literature different types of filters that could be used to obtain a (complex) signal from a real signal to study the vortices [5, 6]. In the case of speckle these filters transform the intensity pattern $I(x,y)$ into a pattern of complex numbers that we call $\tilde{I}(x,y)$.

$$I(x,y) \rightarrow \tilde{I}(x,y) = |\tilde{I}(x,y)|e^{i\theta(x,y)}$$

where $\theta(x,y)$ is called the pseudophase to distinguish it of the actual phase of the acoustical or optical field that constitutes the speckle. Actual phase is not measurable in optics. It is this pseudophase that contains the precise location of vortices where the magnitude of the pseudofield is null and the pseudophase is not defined there. Several types of transform have been proposed in the literature for metrological purposes [7, 8]. We are going to use Laguerre Gauss Transform in its original form and in a generalization obtained by variations the shape of the kernel.

The Laguerre Gauss Transform is defined as [2]:

$$\tilde{I}(x,y) = \iint_{-\infty}^{\infty} LG(f_x, f_y) G(f_x, f_y) e^{2\pi i (f_x x + f_y y)} df_x df_y \tag{1}$$

where $G(f_x, f_y)$ is the Fourier Transform of the image and the Laguerre Gauss (circular) filter $LG(f_x, f_y)$ is:

$$LG(f_x, f_y) = (f_x + if_y) e^{-\left(\frac{f_x^2 + f_y^2}{\omega^2}\right)} \tag{2}$$

with ω the spectral bandwidth of the circular Gaussian filter. It is, the Fourier plane of the object is filtered by a Gaussian band pass filter and multiplied by a linear complex one in the shape of a spiral phase.

**Circular Gaussian kernels**

In our proposal, each intensity image (see Fig. 1 a) is processed by using the LGT to produce a 2D pseudophase map (see Fig. 1 b) containing the vortices. Randomly distributed phase singularities are obtained associated with the speckle pattern that imprints unique marks related to the surface of the object originating it. Notice that the phase singularity positions depend on the parameter ω of the LG transform. To locate the vortices, a modified version of the Goldstein residues method [9] is employed followed by the conventional procedure based on the zero crossing analysis for the real and the imaginary components of the analytical signal. The tracking of homologous vortices is carried out through the evaluation of the topological charge and the corresponding core structure properties, i.e. the vorticity, eccentricity and zero crossing angles [2]. This procedure has been demonstrated to be a powerful tool in precisely determining displacements and low activity in dynamic speckles [4, 10].

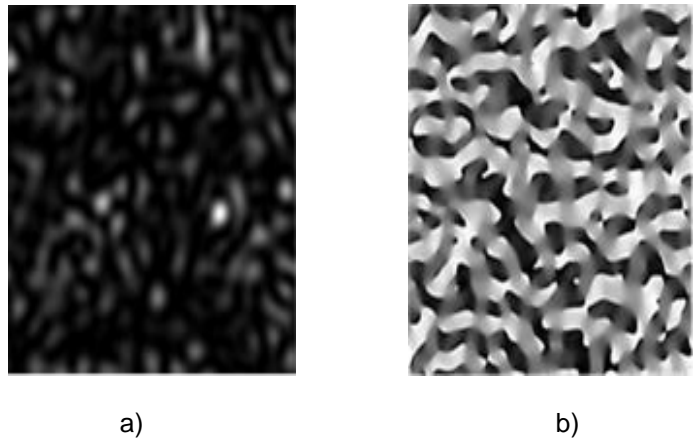

a)          b)

**Fig. 1**. a) Intensity speckle image. b) Pseudo phase map

Usually the Gaussian kernel used in the vortices measurements is chosen circularly symmetric, that is, with a single parameter: Gaussian kernel width ω. This parameter can be slightly varied to obtain a family of very close vortices by including a wider or smaller region of the Fourier space. These varieties of vortices are, in their own right, also attached to points in the object. If the changes in the value of ω are very small, the vortices obtained in this way define a trajectory in every speckled frame (see Fig. 2).

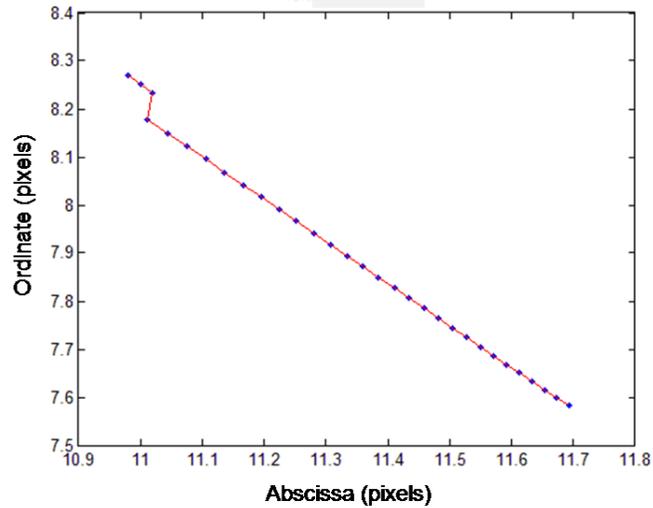

**Fig. 2**. Trajectory of a vortex in an ultrasound image speckle frame obtained by changing the value of the bandwidth ω from 12 to 15 in Fourier space.

In Fig. 2, it is possible to observe the locations of the phase singularities corresponding to different vortices obtained by changing the parameter ω corresponding to a small region around a point in a single frame in an ultrasound (US) image of the heart, basal region of septum. By changing only the ω parameter single trajectories are obtained in each frame of a series. In the next frame of a US series, the homologous speckle describes another trajectory (see Fig. 3).

Vortices are created and/or annihilated or come out of the observation plane in unpredictable ways. So, for the measurements of the trajectory of a region along time it is convenient to verify previously that the chosen vortex remains present in the whole set of images.

By comparing, both qualitatively and quantitatively, the lines corresponding to homologous initial vortices in successive frames, it is possible to extract some conclusions about the movement that occurred in a very small segment of the object. Notice a rigid body translation, a rotation and a step deformation (upper extreme of the red and blue lines).

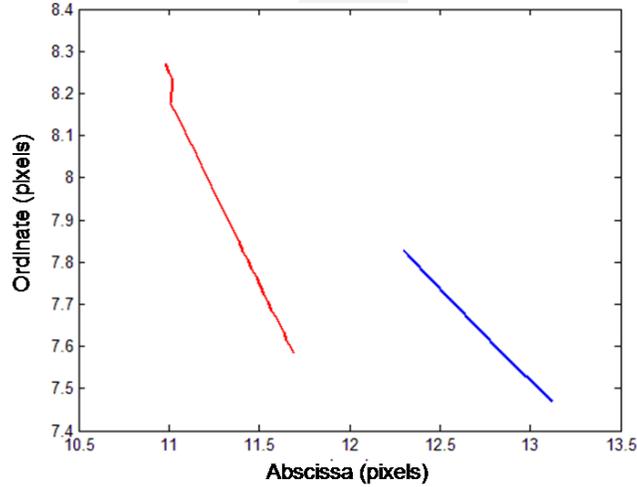

**Fig. 3**. First frame trajectory as in figure 1, repeated in red, and its homologous in the following frame in blue for the same values of the ω parameters.

**Laguerre Gauss Transform (LGT) using elliptical cores**

A similar procedure can be applied by permitting a small generalization of the LGT.

If the kernel is chosen with elliptical geometry, it is defining the Laguerre Gaussian kernel as:

$$LG(f_x, f_y) = (f_x + if_y)e^{-\left(\frac{f_x^2}{\omega_1^2}+\frac{f_y^2}{\omega_2^2}\right)} \quad (3)$$

with two different bandwidths $\omega_1$ and $\omega_2$. The circular core is re obtained when $\omega_1 = \omega_2 = \omega$. In this way we can independently change two widths of the filter, thus including in the calculation slightly different regions of the Fourier plane. The adequate selection of the elliptical kernels parameters allows to obtain a family of vortices in the neighborhood of a given one in arbitrarily close locations. The amplitude corresponding to an elliptical kernel is shown in Fig. 4.

The $\omega_1$ and $\omega_2$ parameters permit to control the bandwidth of the Laguerre Gauss kernel in each axis by including into their band passes additional regions of the Fourier plane in different regions. Of course, this expression shows the elliptical core in its canonical representation. It can be further generalized to include rotated elliptical kernels.

We show now some possibilities provided by this generalization: annihilation of opposite charged vortices can be located, identification of homologous vortices in different frames improvement, rigid body movements (displacements and rotations) and the possibility of the measurement of very localized strains.

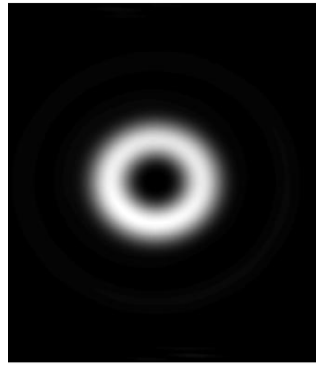

**Fig. 4**: a) Amplitude of the elliptical kernel.

More than a set of vortices can be obtained intersected in a single point by variation of both parameters in different ways (see Fig 5).

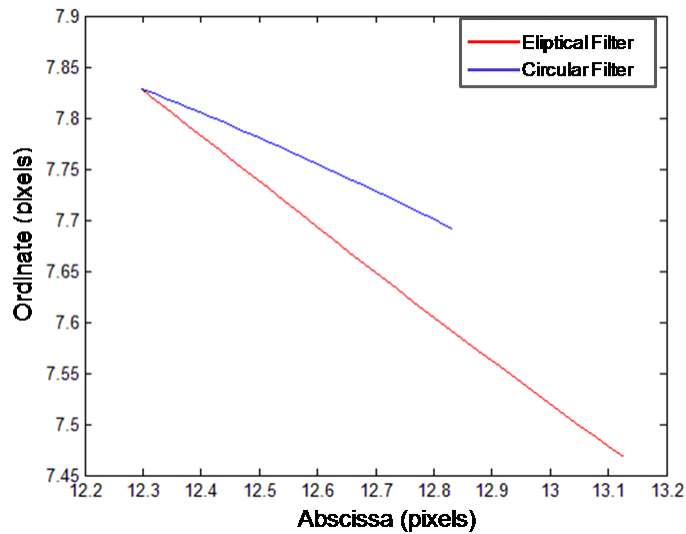

**Fig 5**: Two families of vortices obtained from a single frame and changing the parameters in two different ways (blue curve correspond to the LGT with circular kernel and red curve corresponds to the LGT with elliptical kernel).

## Annihilation of opposite charged vortices in a single frame

In optical vortex metrology, the changes in vortex networks in a dynamic event analysis include the creation and annihilation of pairs of vortices with opposite topological charge. Vortices usually annihilate by pairs of opposite topological charge. In a single frame, we can find the lines obtained by omegas changes and estimate where the pair is going to annihilate by varying the omega parameters.

Now we analyze the annihilation of such pairs of vortices when the $\omega_1$ and $\omega_2$ parameters in the LG elliptical kernel are slightly changed in the same frame. The experimental arrangement used for these results is schematized in Fig. 6. A collimated laser beam illuminates a random diffuser with a 5 mm pupil aperture. The distance between the diffuser and the camera plane is 50 cm. In our proposal, we record two speckled images corresponding to the states before (Fig. 7) and after a diffuser displacement in the horizontal x-axis, $\Delta x$=60 mm. The illuminated area onto the diffuser slightly changes, and consequently, a rather different set of radiators is associated with waves interfering at the camera plane. Then, the speckle changes producing slight decorrelation between images. We calculated the pseudophase from the complex analytic signal of the 2-D speckle pattern by applying the elliptical LG filter. The oppositely charged vortices pairs are indicated in the green box in Fig. 7 a) and 7 b). The vortices with positive and negative topological charge are indicated in blue and red respectively. As the parameter $\omega_1$ in the elliptical LG transform slightly changed, the vortices in the 2D pseudo phase-maps displace to near positions. The vortices locations in connection with LG filter change are observed in Fig. 7 a) and 7 b) for both opposite topological charge vortices.

The exact point where annihilation happens is somehow difficult to find but it can be almost indefinitely approximated. The change in the omega values required for this approximation becomes progressively smaller as the two vortices approach.

The vortices locations in connection with LG filter change by variation of the omega parameters in different ways are observed in Fig. 8 for both opposite topological charge vortices.

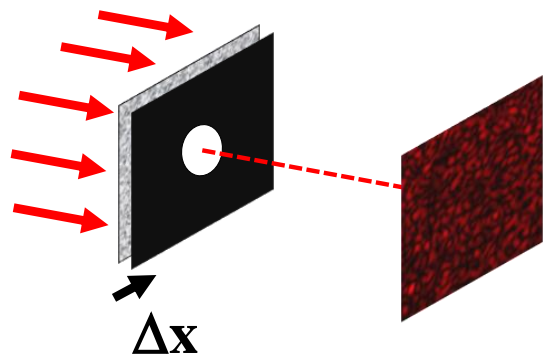

**Fig. 6**. Experimental set up.

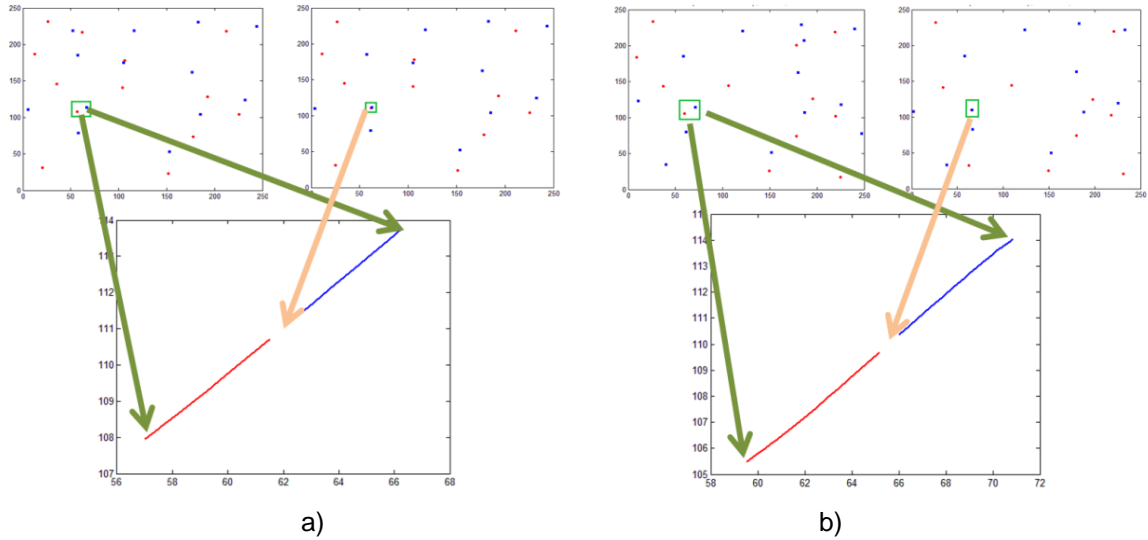

a)  b)

**Fig. 7**

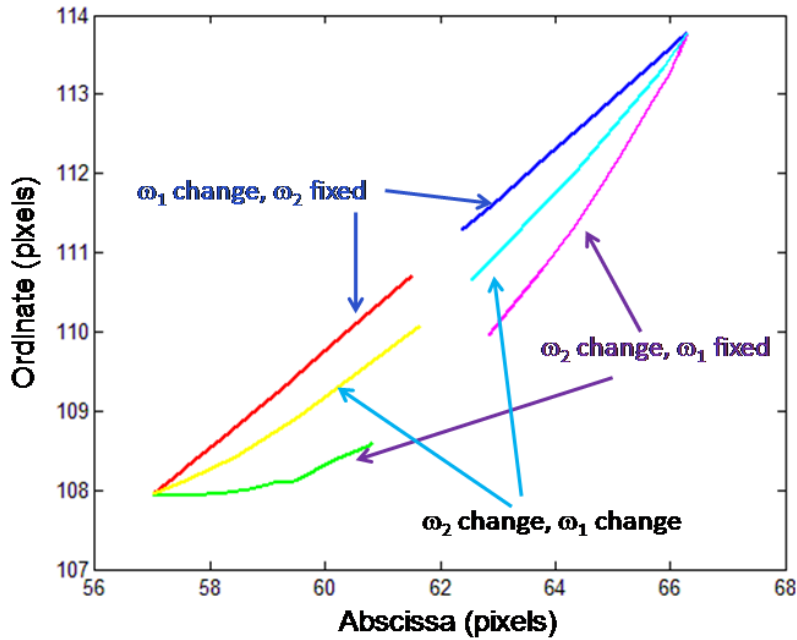

**Fig. 8**. Approaching the annihilation point in several different trajectories obtained by changing the omega values in linear combinations.

**Clouds and local average (local) strain: Application to the measurement of cardiac strain**

Small variations on the kernel parameters ($\omega_1$, $\omega_2$) add small regions of the Fourier plane and slightly change the location of the vortices. Each variation can be considered as a vortex in its own right and used to track its displacement looking for the position of its homologous (with the same $\omega_1$, $\omega_2$ values). So, a number of close vortices can be created around a first one by changing the $\omega_1$, $\omega_2$ in different ways. We are going to call a *cloud* to such set.

By considering two vortices in the same frame and their associated clouds, several measurements of the strain can be done in quite similar conditions between vortices that are localized at very short distances.

Every cloud is characterized by a mass center of its vortices and a dispersion of its positions.

When the strain is calculated between homologous vortices in both clouds in consecutive frames, a mean value of the strain and the standard error of its mean can be obtained.

The measurement of strain is extremely important in echocardiography. The results obtained, usually by speckle tracking method [11-13] are, generally pooled in a single curve involving the average of the strains in different regions of the heart in what is called Global Longitudinal Strain provide clinical indications concerning.

Nevertheless, the possibility to measure strain on a very local basis is recognized as an interesting possibility.

**Strain Cardiac Curve**

We analyze now the trajectory of a pair of vortex clouds in a series of US images along a full cardiac cycle of a human healthy patient and calculate the local strain in every frame.

By using the clouds vortex method we found a case with an interesting phenomenon worth to be taken into account.

Strain, a dimension less magnitude usually expressed as a percentage, is defined as usual as [10]:

$$\varepsilon \equiv \frac{L - L_0}{L_0}$$

where $L_0$ is the length of a segment of an object considered as reference and $L$ is the length of its homologous segment in its strained state.

It takes two clouds in each frame to calculate strains. The reference state $L_0$ is the distance between the clouds chosen as usual in the literature at the point R of ECG.

When the strains are calculated using all the values of the clouds the strain curve gives the result shown in Fig. 10.

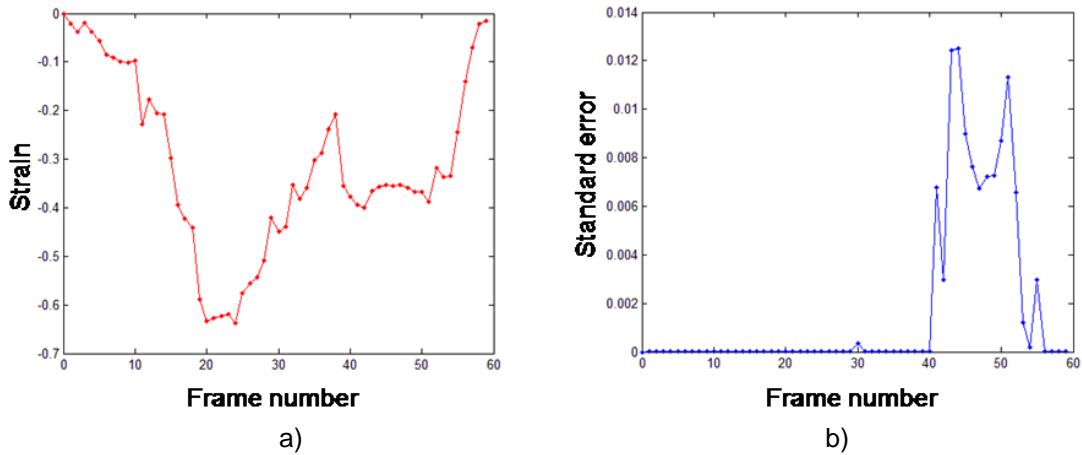

a)                                               b)

**Fig. 10** a) The mean value of the local strain using 25 samples. b) The standard error of the mean. While the standard error of the mean is negligible small in the right part (up to frame 39), a sudden significant increase happens there up to about frame 55.

A careful observation of the individual averaged strains shows that there are two sets of curves with different behavior (see Fig. 11). Notice the branching in the GLS curve shown as the red and blue arms after frame 40 and merging about 55. They differ up to about a 10% in the branch region and very little in the rest of the cycle.

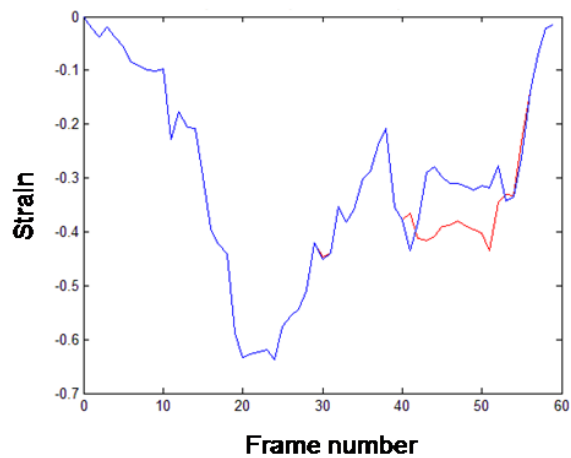

**Fig. 11.** The strain curve with 15 samples follow the blue curve. The strain curve with 10 samples follows the blue line up to frame 40 and the red one after that and merges with the blue one at about frame 55.

If both branches are shown separately we observe the red plot of Fig. 12 and Fig.13.

The mean strain curve of Fig. 12 a) and Fig. 13 a) corresponds to the red and blue plot in Fig. 11), respectively. Curves in Fig. 12 b) and 13 b) shows the standard error of the mean. Notice that the error curve is smaller than 4.2 times (see Fig 12 b))and than$10^{-5}$ 5.5 times $10^{-5}$ (see Fig. 13 b) It is considerable smaller than the results obtaining current Speckle Tracking methods.

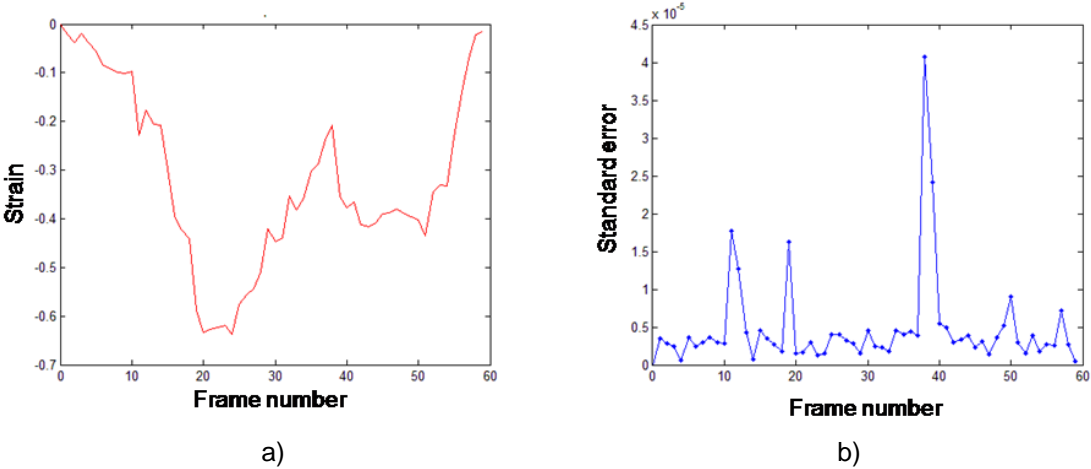

a)                            b)

**Fig. 12**. a) Mean strain curve (corresponds to the red branch in Fig.11). b) Standard error of the mean (notice the scale factor $10^{-5}$)

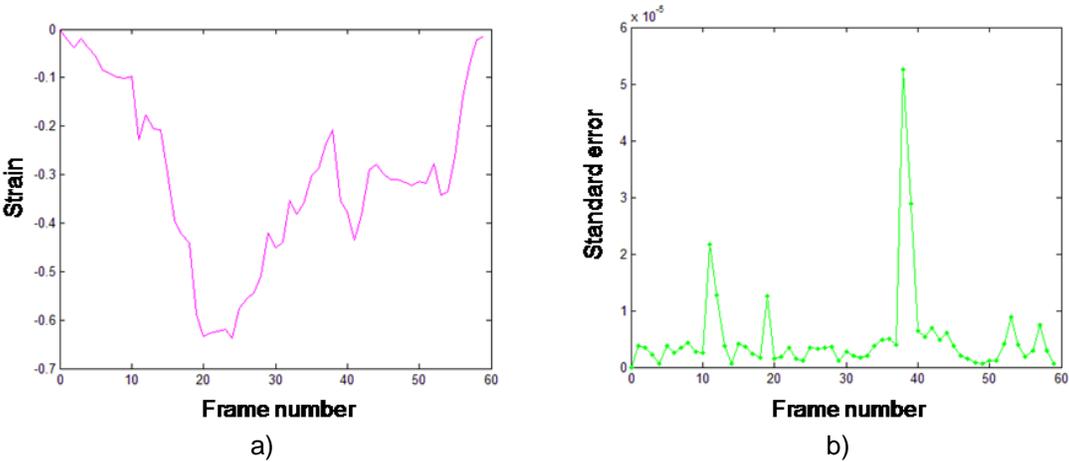

a)                            b)

**Fig. 13**: a) Mean strain curve (corresponds to the blue plot in Fig. 11) b) Standard error of the mean (notice the scale factor $10^{-5}$)

The maximum error is committed near the branching and is significantly smaller for the rest of the curve.

Both trajectories show a very small standard error of the mean. Very much smaller than the separation in the branching, thus showing that the difference cannot be attributed to chance.

The Student t test provides the probability that such a chance departure is extremely low (see Fig. 14).

Not so outstanding, there seem to exist a reflected similarity between both branches, the red occurring a few frames before the blue one.

A similar behavior was found in other strain curves. We estimate that these local and precise measurements cannot be resolved with current speckle tracking techniques.

The probability α that these results were obtained by chance (it is, the nonexistence of the bifurcation) is negligibly small (less than 0.0005)

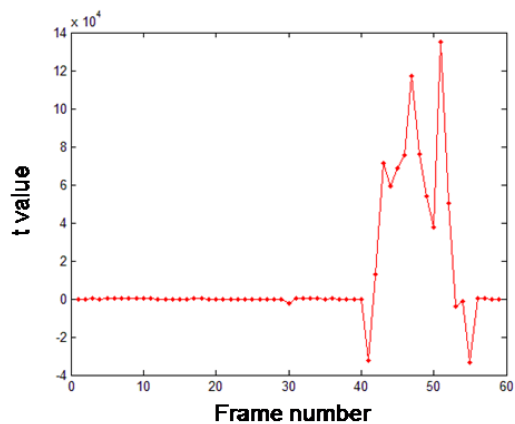

Fig. 14

**Conclusions**

We have proposed the variation of the circular Gaussian kernel bandwidth and a modification of the Laguerre Gauss kernel in Laguerre Gauss Transform to identify vortices of the pseudo field with very close locations. The last consists in the use of elliptical gaussian kernels with slightly different ellipticities. It provides an interesting tool in its use as a metrological tool. Rigid body translation and rotation can be readily identified. Deformations also appear in a visual way. We explore how pairs of such vortices annihilate.

Cloud measurement of the average strain showed that the local strain curve had a bifurcation that could not be detected with ST and that the standard error of the mean in this case was considerably smaller. Even if we cannot evaluate the eventual clinic use (or value) of these results, they show that vortices clouds give results with low standard error of the mean and could be considered for further study.

The estimated distance between the vortices obtained in ultrasound images by slight changes of the omega parameters in a cloud is about 0.5 μm. A cardiac muscular fiber is about 15 μm width, it is, some 30 times bigger. This indicates that vortex metrology could, in principle, be more precise than what is required to resolve such fibers.

Localized abnormalities could be detected and quantified if the pattern of deformation in a given heart were compared to the normal range for that region because normal motion and strain in the left ventricle is spatially heterogeneous.

Further technical development and standardization of methodology are necessary before clinical applications can be recommended.

The concept of vortex clouds obtained by using elliptical cores could also be applied to dynamic speckle activity measurements but it is going to be treated in a further paper.

**Acknowledgements**

This work was supported by Consejo Nacional de Investigaciones Científicas y Técnicas PIP 849 and 2087, by Facultad de Ingeniería, University of La Plata Grant, ANPCyT Grant Nº. 2087 and 4558 and Comisión de Investigaciones Científicas de la Provincia de Buenos Aires, Argentina. We thank Dr. Jorge Lowenstein for providing the US images and very helpful advice.